\documentclass[11pt]{article}

\usepackage{amsmath}
\usepackage{graphicx}
\usepackage{indentfirst}
\usepackage{amssymb}
\usepackage{cite}
\usepackage{color}
\usepackage{subfigure}

\begin{document}

\title{Violations of the weak cosmic censorship conjecture in the higher dimensional $f(R)$ black holes with pressure}

\date{}
\maketitle

\begin{center}
\author{Ke-Jian He,}\footnote{kjhe94@163.com}
\author{Guo-Ping Li,} \footnote{gplipys@yeah.net}
\author{Xin-Yun Hu}$^{*}$\footnote{Corresponding author: huxinyun@126.com}\\

\vskip 0.25in
$^{1}$\it{Physics and Space College, China West Normal University, Nanchong 637000, China}\\
$^{2}$\it{Physics and Space College, China West Normal University, Nanchong 637000, China}\\
$^{3}$\it{College of Economic and Management, Chongqing Jiaotong University, Chongqing 400074, China}

\end{center}
\vskip 0.6in
{\abstract
{
We adopt the energy  momentum relation of  charged particles to study the thermodynamics laws and weak cosmic censorship conjecture of $D$-dimensional $f(R)$ AdS black holes  in different phase spaces by considering charged particle absorption. In the normal phase space,  it turns out that the laws of thermodynamic  and the weak cosmic censorship conjecture are valid. In the extended phase space, though the first law of thermodynamics is  valid, the second law of thermodynamics is invalid. More interestingly, the weak cosmic censorship conjecture is shown to be violated only in higher-dimensional near-extremal $f(R)$ AdS black holes. In addition, the magnitudes of the violations for both  the second law  and weak cosmic censorship conjecture are dependent on  the charge $Q$, constant scalar curvature $f'(R_0)$, AdS radius $l$, dimension parameters $p$, and their variations. }
}

\thispagestyle{empty}
\newpage
\setcounter{page}{1}

\section{Introduction}\label{sec1}
The event horizon is indispensable for a  black hole, since not only both the temperature and entropy are related to it, but also the singularity of the black hole should be completely obscured by the event horizon. If the singularity of the black hole  exposed or the event horizon is destroyed, the bare singularity will destroy the causal relationship in the spacetime. In order to avoid the occurrence of this phenomenon, Penrose  proposed the weak cosmic censorship conjecture\cite{ref1,ref2}, which supposed that the singularity of the black hole was always hidden by the event horizon. Though the weak cosmic censorship conjecture seems to be reasonable, there is no general method to prove the correctness of this conjecture so far. Hence, it is necessary  to test the validity of this conjecture for different types of black holes. An alternative thought procedure was developed by Wald to test the stability of event horizons of black holes interacting with test particles or fields \cite{ref3}. Based on this idea, it was found that the particle with sufficient charge and angular momentum would not be absorbed by the extremal Kerr-Newman black hole. In other words, the event horizon of the extremal Kerr-Newman black hole could not be destroyed by the particle, and the weak cosmic censorship conjecture is still valid. Then, this result was also generalised to scalar field \cite{ref4,ref5}. Nevertheless,  Hubeny  pointed out that the near-extremal Reissner-Nordstr\"{o}m black hole  would be overcharged by absorbing the particle, thereby, the weak cosmic censorship conjecture would be invalid \cite{ref6}. Similarly, the near-extremal Kerr black hole could be overspun, and the event horizon was also unstable \cite{ref7,ref8}. Later,  when the back-reaction and self-force effects were taken into account \cite{ref9,ref10,ref11,ref12,ref13}, the conjecture was found to be valid for the  near-extremal Reissner-Nordstr\"{o}m black hole and the near-extremal Kerr black hole. Hence, the check of the weak cosmic censorship remains one of the most essential open topics in classical general relativity. At present, there have been a lot of studies concentrating on the weak cosmic censorship conjecture in different spacetime \cite{ref14,ref15,ref16,ref17,ref18,ref19,ref20,ref21,ref22,ref23,ref24,ref25,ref26,ref27,ref28,ref29,ref30,ref31,ref32,Wang:2019jzz,ref33,ref34,ref35,ref36,ref37,ref38,ref39,ref40,Chen:2019nsr}.

Recently,  Ref.\cite{ref41} stated that the laws of thermodynamics and weak cosmic censorship conjecture can be tested when the charged particle dropped into the black hole. Based on his work, the first law of thermodynamics of higher-dimensional  Reissner-Nordstr\"{o}m black hole   was found to be  valid under charged particle absorption. Furthermore, they found that the extremal black hole kept the initial state and was not overcharged. Therefore, it was claimed that the weak cosmic censorship conjecture was valid in the extended phase space. However, one can see that  the second law of thermodynamics is not valid under the absorption which would only be seen in the case considering the pressure and volume term. The cosmological constant is a parameter which plays an important role in determining the asymptotic topology of a black hole spacetime, and it  set to be a constant value  in the action of Einstein gravity. In fact, imposing the cosmological constant as a dynamical variable is prevails now. In this case, thermodynamics was studied more widely in the expended phase spaces where the cosmological constant is identified as thermodynamic pressure \cite{Caldarelli:1999xj}, and its conjugate is found to be thermodynamic  volume \cite{Dolan:2010ha,Cvetic:2010jb,Kastor:2009wy}. Soon after, the laws of thermodynamics and weak cosmic censorship conjecture were checked in the Born-Infeld AdS balck holes and phantom Reissner-Nordstr\"{o}m black holes \cite{ref30,ref32} in the different phase spaces. Differently from the study in Ref.\cite{ref41}, they did not employ any approximation, and found that extremal black holes change into non-extremal black holes for the absorbed particle.
Similarly, they also found the violation of the second law of thermodynamics in the expended phase spaces.

Among the researches mentioned  above, the weak cosmic censorship conjecture of higher-dimensional $f(R)$ black holes under charged particle absorption has  not yet been reported. As well as known, the $f(R)$  gravity as a highly valued model of modified general relativity is very important, it may provide a feasible explanation for the accelerated expansion of the universe \cite{ref43,ref44}. When one considers $f(R)$ theory as a modification of general relativity, it is fairly essential to study the features  of black holes in this theory, and the thermodynamics of the black hole is also an essential subject in the theory of gravity. In view of this, various investigations have been discussed with respected to the thermodynamics in the $f(R)$ spacetimes \cite{ref46,ref47,ref48,ref49,Moon:2011hq,Myrzakulov:2015qaa,Guo:2019pzq}. In these studies, they found that the  laws of thermodynamics of $f(R)$  black hole were accurate. Motivated by these facts, our aim is to promote the work of  Ref.\cite{ref41} to the  higher-dimensional $f(R)$ AdS black hole,  where a more accurate calculation is presented. We will use the test particle model to study the thermodynamic laws and weak cosmic censorship conjecture of the higher-dimensional $f(R)$ black holes. What's more,  we will also explore  whether  $f(R)$ gravity parameters will affect the second law and weak cosmic censorship conjecture. As a result, we find that the first law is still valid  in   different phase spaces, and the extremal black holes  are still extremal after an absorption of the external particle. However, the second law is violated in the extended phase space though it is valid in the normal phase space. More importantly, we also find  that the weak cosmic censorship conjecture is valid under the case of without pressure, while for the case with pressure, the weak cosmic censorship conjecture is violable, depending on $f'(R_0)$ gravitational parameters.

The remainder of this article is organized as follows. In section 2, we introduce higher-dimensional $f(R)$ AdS black holes and its first law of thermodynamics. In section 3, the motion  of charged particle in higher-dimensional $f(R)$ AdS black holes is  investigated. In section 4, the laws of  thermodynamics of higher-dimensional $f(R)$ AdS black holes are checked in the different phase spaces. In section 5, the validity of  the weak cosmic censorship conjecture in different phase spaces are checked with a more accurately examine. In section 6, we briefly summarize our results.  In this paper, we will set $G = c= 1$.

\section{A brief review on the higher-dimensional  $f(R)$ black holes} \label{sec:2}
 Except for  the simple and general Lagrangian model, $f(R)$ gravity also take into account arbitrary function of Ricci scalar. However, the standard Maxwell energy-momentum tensor is not traceless in higher dimensions. Hence, it is important to  find that the higher-dimensional black hole solutions in $R+f(R)$ gravity coupled to standard Maxwell field. In general, the conformally invariant Maxwell action  in arbitrary dimensions  is  given by \cite{ref50}
\begin{equation}
S_m=-\int d^Dx\sqrt{g}\left(F_{\mu \nu }F^{\mu \nu }\right){}^p,\label{metric1}
\end{equation}
in which $p$ is a positive integer, i.e, $p\in\mathbb{ N}$. $F_{\mu \nu }=\partial _{\mu }A_{\nu }-\partial _{\nu }A_{\mu }$  is the electromagnetic tensor, where $A_{\mu } $ stands for  the electromagnetic potential.  It can be evidenced that  the energy momentum tensor  is traceless  when $D = 4p$.  In the special case  $p = 1$, the above equation is reduced to the standard Maxwell action. Therefore, the action of $R+f(R)$ gravity in $D$-dimensional spacetime coupled to a  conformally invariant  Maxwell field reads
\begin{equation}
S=\int _\mathcal{M} d^Dx\sqrt{g}\left[R+f(R)-\left(F_{\mu \nu }F^{\mu \nu }\right){}^p\right],\label{metric2}
\end{equation}
where $f(R)$ is an arbitrary function of scalar curvature $R$. Then, $D$-dimensional black hole metric is described as follow \cite{ref46}
\begin{equation}
{ds}^2=-W(r)d t^2+\frac{{dr}^2}{W(r)}+r^2 {d\Omega }_{D-2}^2,\label{metric3}
\end{equation}
and
\begin{equation}
W(r)=1-\frac{2m}{r^{D-3}}+\frac{q^2}{r^{D-2}}\times \frac{\left(-2q^2\right)^{(D-4)/4}}{\left(1+f'\left(R_0\right)\right)}-\frac{R_0r^2}{D(D-1)}.\label{metric4}
\end{equation}
It is important to note that  the above black hole solutions hold for the dimensions which are multiples of four, since the assumption of traceless energy-momentum tensor is crucial for deriving an accurate solution of the black hole in the gravitational force of $f(R)$  coupled to the matter field. Hence, the solution exist only for $D = 4p$ dimensions.  In order to have a real solution we should  follow the restriction $D = 4p$, i.e., $D = 4,8,12,....,$ which means that $p$ should be only a positive integer \cite{ref46}.
In accordance with  Ref.\cite{ref46}, the above solution is asymptotically AdS when $R_0=-D(D-1)/l^2$.
In addition, the parameters $m$ and $q$  are  integration constants which are related to the mass $M$ and electric charge $Q$,  and we have  \cite{ref46}
\begin{equation}
M=\frac{(D-2)\Omega _{D-2}}{8\pi }m\left(1+f'\left(R_0\right)\right),  \label{eq2.5}
\end{equation}
\begin{align}
Q=\frac{D(-2)^{(D-4)/4}q^{(D-2)/2}\Omega _{D-2}}{16\pi \sqrt{1+f'\left(R_0\right)}}.  \label{eq2.6}
\end{align}
As one can see from equation (\ref{metric4}), the solution is ill-defined for $f'\left(R_0\right) = -1$. In the other hands, there would be inner and outer horizons and an extreme black hole or naked singularity due to different choices of parameters when $1+f'\left(R_0\right)>0$. However, for  the case $1+f'\left(R_0\right)<0$, the conserved quantities such as mass would be negative, making this case nonphysical,  thus this is not a physical case and we do not consider this situation\cite{ref46,ref48}.
At the outer event horizon $r = r_h$, the Hawking temperature $T_h$,  entropy $S_h$, and electric potential $\Phi_h$ are obtained as \cite{ref48}
\begin{align}
&T_h=\frac{1}{4\pi }\left(\frac{\partial W(r_h)}{\partial r_h}\right) \nonumber\\
&~~~=\frac{\left(\left(1+f'\left(R_0\right)\right)\times \left[2r_h{}^2(D-1)+2 l^2(D-3)\right]+\left(-2q^2\right)^{D/4}r_h{}^{2-D} l^2\right)}{8 l^2 \pi  r_h\left(1+f'\left(R_0\right)\right) },\label{eq2.7}
\end{align}
\begin{align}
S_h=\int _0^{r_h}T^{-1}\left(\frac{\partial M}{\partial {r_h}}\right)_{Q}{d{r_h}}=\frac{r_h{}^{D-2}\Omega _{D-2}}{4}\left(1+f'\left(R_0\right)\right),  \label{eq2.8}
\end{align}
\begin{align}
\Phi_h =\frac{q}{r_h}\sqrt{1+f'\left(R_0\right)},\label{eq2.9}
\end{align}
where $\Omega _{D-2}$ denotes the volume of the unit $(D-2)$-sphere.
Therefore, the first law of thermodynamics at the cosmological horizon is expressed as \cite{ref48}
\begin{align}
dM=T_{h}dS_{h}+\Phi_{h}dQ.\label {eq2.10}
\end{align}
In the extended thermodynamic phase space, the cosmological constant is identified as the thermodynamic pressure while its conjugate quantity is regarded as the thermodynamic volume. The pressure is defined as
\begin{align}
P=\frac{-\Lambda }{8\pi }=\frac{(D-1)(D-2)}{16{\pi l}^2}.  \label{eq2.11}
\end{align}
The expression of the cosmological constant in the $D$-dimensional spacetime  is $\Lambda =-\frac{(D-1)(D-2)}{2l^2}$, where  $l$ is the radius of the AdS space.
Therefore,  the relation between $\Lambda$ and $R_0$ is $R_0=\frac{2D \Lambda }{D-2}$, which can be reduced to the relation $R_0 = 4\Lambda$ when $D = 4$.
Based on equations (\ref{metric4}), (\ref{eq2.5}) and (\ref{eq2.11}), one can derive
\begin{equation}
V_h=\left(\frac{\partial M}{\partial P}\right)_{S,Q}=\frac{\left(1+f'\left(R_0\right)\right){  }r_h{}^{D-1}\Omega _{D-2}}{D-1}.\label{eq2.12}
\end{equation}
Due to the effect of $f(R)$ gravity, the expression of equation (\ref{eq2.12}) incldueds an extra factor $1 + f¡ä(R_0)$. What is more, these thermodynamic quantities obeys the first law of thermodynamics
in the extended phase space, which is \cite{ref48}
\begin{align}
dM=T_{h}dS_{h}+\Phi_{h}dQ+V_{h}dP.  \label{eq2.13}
\end{align}
And, the following Smarr relation is also satisfied
\begin{equation}
M=\frac{D-2}{D-3}T_h S_h+\frac{(D-2)^2}{D(D-3)}\Phi_h  Q-\frac{2}{D-3}V_h P. \label{eq2.14}
\end{equation}
In the extended phase space, the mass of black holes should be interpreted as enthalpy. The relation among the enthalpy, internal energy and  pressure  is
\begin{align}
M=U_h+PV_h,  \label{eq2.15}
\end{align}
where $U_h$ is internal energy. Hence, the change of the mass makes re-balance not only for the horizon and  electric charge, but also the AdS radius in $PV_h$ term.

\section{Charged particle absorption in higher dimensional charged $f(R)$ black holes}\label{sec3}

In this section, we are going to consider the dynamic behavior of the charged particle which is  near the event horizon, and we focus on the dynamic behavior of a charged particle swallowed by the black hole and its energy-momentum relationship. In this process, we are mainly consider  the scalar particle, and  the motion of scattered particles satisfy the Hamilton-Jacobi equation of curved spacetime, which is
\begin{align}
g^{\mu \nu }\left(p_{\mu }-{eA}_{\mu }\right)\left(p_{\nu }-{eA}_{\nu }\right)+{m_b} ^2=0,  \label{eq3.1}
\end{align}
and
\begin{align}
 p_\mu=\partial_\mu \mathcal{I},  \label{eq3.2}
\end{align}
where $m_b$ and $e$ are the rest mass and charge of the particle respectively, $p_\mu$ is the momentum of the particle, and $\mathcal{I}$ is the Hamilton action of the particle. Taking into account the symmetries of the spacetime,  the Hamilton action of the moving particle can be separated into
\begin{align}
\mathcal{I}=-{\omega t}+I_r(r)+\underset{i=1}{\overset{d-3}{\sum }}I_{\theta_i }(\theta_i )+{L\psi },  \label{eq3.3}
\end{align}
in which the conserved quantities  $\omega$ and $L$  are  energy and angular momentum of particle, which  are assumed from the translation symmetries of the metric in equation (\ref {metric3}), and they are conserved quantities of spacetime in the gravitational system. In addition, $I_r(r)$ and $I_{\theta_i }(\theta_i )$ are the  radial-directional component and ${\theta}$-directional component of the  action respectively.  Owing to $D$-dimensional solution, the black hole includes a $D-2$-dimensional sphere $\Omega _{D-2}$, and the angular momentum $L$ corresponding to the translation symmetry of the last angle coordinate of $\Omega _{D-2}$.
Then, the $(D-2)$-dimensional sphere can be written as
\begin{align}
h_{{ij}}{dx}^i{dx}^j=\underset{i=1}{\overset{D-2}{\Sigma }}\left(\underset{j=1}{\overset{i}{\Pi }}\sin ^2\theta _{j-1}\right){d\theta }_i^2, \quad \theta _{D-2}\equiv \psi.  \label{Eqspherically}
\end{align}
To solve the Hamilton-Jacobi equation,  we can use the contravariant metric of the black hole, with help of equation (\ref{metric3}),  we  obtain
\begin{align}
g^{\mu \nu }\partial _{\mu }\partial _{\nu }=-W(r)^{-1}\left(\partial _t\right){}^2+W(r)\left(\partial _r\right){}^2+r^{-2}\underset{i=1}{\overset{D-2}{\Sigma }}\left(\underset{j=1}{\overset{i}{\Pi }}\sin ^{-2}\theta _{j-1}\right)\left(\partial _{\theta_i }\right){}^2.  \label{eq3.4}
\end{align}
Substituting above equations into equation (\ref{eq3.1}), the Hamilton-Jacobi equation can be re-expressed as
\begin{align}
&-{m_b}^2=-\frac{1}{W(r)}\left(-\omega-{eA}_t\right){}^2+W(r)\left(\partial _rI(r)\right){}^2+r^{-2}\underset{i=1}{\overset{D-3}{\Sigma }}\left(\underset{j=1}{\overset{i}{\Pi }}\sin ^{-2}\theta _{j-1}\right)\left(\partial _{\theta_i }I(\theta_i )\right){}^2\nonumber\\
&~~~~~~~~~+r^{-2}\left(\underset{j=1}{\overset{D-2}{\Pi }}\sin ^{-2}\theta _{j-1}\right)L^2.  \label{eq3.5}
\end{align}
We can  separate equation (\ref{eq3.5}) by introducing a variable $\mathcal{R}$. Therefore, the radial and angular components are
\begin{align}
-\frac{r^2}{W (r)} \left(-\omega -{eA}_t\right){}^2+r^2W (r) \left(\partial _rI(r)\right){}^2+r^2{m_b} ^2=-\mathcal{R},  \label{eq3.6}
\end{align}
\begin{align}
\underset{i=1}{\overset{D-3}{\Sigma }}\left(\underset{j=1}{\overset{i}{\Pi }}\sin ^{-2}\theta _{j-1}\right)\left(\partial _{\theta_i }I(\theta_i )\right){}^2+\left(\underset{j=1}{\overset{D-2}{\Pi }}\sin ^{-2}\theta _{j-1}\right)L^2=\mathcal{R}.  \label{eq3.7}
\end{align}
The radial- and $\theta$-directional equations are sufficient to obtain the relation between energy and electric charge of the particle.  In addition, we are interested only in the location at the horizon. Concretely, we pay attention to  the radial momentum of the particle, that is, we consider angular moment $L$ is a constant depending on the trajectory of the particle. Furthermore, we obtain the radial momentum
\begin{align}
p^r\equiv g^{rr}\partial_r I(r)=W(r)\sqrt{\frac{-{m_b} ^2r^2-\mathcal{R}}{r^2W(r)}+\frac{1}{W(r)^2}\left(-\omega -{eA}_t\right){}^2}.\label{eq3.8}
\end{align}
As $\mathcal{R}$ is eliminated, near the event horizon where $W(r) \rightarrow 0$, the above equation is reduced to
\begin{align}
\omega=\Phi_h e+ p^r_h.  \label{eq3.9}
\end{align}
Here, equation (\ref{eq3.9}) is the relation between conserved quantities and momenta for a given radial location $r_h$. For a special case $\omega=\Phi_h e$, the energy of the black hole does not change. However, for the case $\omega<\Phi_h e$, the energy of the black hole flows out the horizon,  which leads to the supperradiation occurs \cite{ref33}. Therefore, it is stressed that a positive sign should be endowed in front of $ p^r_h$  in order to assure a positive time direction, which means we should chose $\omega\geq\Phi_h e$  \cite{ref51}. Hence, in the positive direction of time,  the energy and momentum of the particle are positive.

\section{Thermodynamic  of the higher dimensional  $f(R)$ black holes under charged particle absorption}
\label{sec:4}
Black holes can be viewed as thermodynamic systems since they do not only have temperature and entropy, but also energy and chemical potential.  In Refs. \cite{ref46,ref48}, thermodynamics of higher-dimensional $f(R)$  AdS black holes in different phase spaces  have  been reported.  In this section,  we would like to generalize those research to the thermodynamics under charged particle absorption. Absorbing a charged particle, the higher dimensional charged $f(R$) black hole is varied by the same quantity as that of the particle, and the variations of the black  hole  energy and charge can be calculated. Subsequently, we can further study the validity of the laws of thermodynamics in different phase spaces under the particle absorption.
\subsection{Thermodynamics in the normal phase space }

In the process of absorption, the energy and electric charge  of a particle are equal  to the change of  internal energy and charge of the black hole. In the normal phase space,  the mass was interpreted as internal energy, that is
\begin{align}
\omega=dM,   e=dQ.  \label{eq4.1}
\end{align}
Using this relation, the energy momentum relation in equation (\ref{eq3.9}) can be expressed as
\begin{align}
dM=\frac{q}{r_h}\sqrt{1+f'\left(R_0\right)} dQ+p^r_h. \label{eq4.2}
\end{align}
Obviously, we need find the variation of entropy in order to rewrite equation (\ref{eq4.2}) to the first law of thermodynamics. Therefore, in accordance with equation (\ref{eq2.8}), as the charged particle is absorbed by the black hole,  the variation of entropy can be written as
\begin{align}
{dS}_h=\frac{1}{4} (D-2){  }\left(1+f'\left(R_0\right)\right)r_h{}^{D-3} \Omega _{D-2} {dr}_h,  \label{eq4.3}
\end{align}
where $ dr_h$ is the variation of  event horizon of the black hole. The event horizon  changes as it absorbs a  particle, and this give rise to a change of $W(r)$.  Thus, the change of  $d{W_h}$ satisfy
\begin{align}
d{W_h}=\frac{\partial W_h}{\partial M}{dM}+\frac{\partial W_h}{\partial Q}{dQ}+\frac{\partial W_h}{\partial r_h}{dr}_h=0,\quad W_h=W\left(M,Q,r_h\right).  \label{eq4.4}
\end{align}
In the normal phase space, the cosmological constant  is fixed. The initial state of  black hole is represented by $(M,Q,r_h)$,
where
\begin{align}
&\frac{\partial W_h}{\partial M}=-\frac{16\text{  }\pi  {r_h}^{3-D}}{(D-2) \left(1+f'\left(R_0\right)\right)\Omega _{D-2}}{dM},  \quad \nonumber\\
&\frac{\partial W_h}{\partial Q}=-\frac{2^{\frac{D}{4}-1} D{  }r_h{}^{2-D}{  }\mathcal{A}}{(D-2) \left(1+f'\left(R_0\right)\right) Q}{dQ},\quad \nonumber\\
&\frac{\partial W_h}{\partial r_h}=\frac{1}{2} r_h \left(\frac{4}{l^2}+\frac{r_h{}^{-D} \left(32 (D-3) M \pi  r_h+2^{\frac{d}{4}} (D-2)^2 \mathcal{A} \Omega _{D-2}\right)}{(D-2) \left(1+f'\left(R_0\right)\right) \Omega _{D-2}}\right){dr}_h,  \label{eq4.5}
\end{align}
and
\begin{align}
\mathcal{A}=\left(-\pi ^{\frac{4}{D-2}} \left(-\frac{(-1)^{-\frac{D}{4}} 2^{5-\frac{D}{4}} \sqrt{1+f'\left(R_0\right)} Q}{D{  }\Omega _{D-2}}\right){}^{\frac{4}{D-2}}\right){}^{D/4}. \label{eq4.6}
\end{align}
Combining equations (\ref{eq4.2}) and (\ref{eq4.4}), $dM$ and $dQ$ will be removed, and we can get $dr_h$ directly
\begin{align}
{dr}_h=\frac{(D-2)^{-1}l^2r_h{}^3 \left(32 \pi  p^r_h r_h+32 \sqrt{1+f'\left(R_0\right)} \mathcal{B} {dQ} +2^{\frac{D}{4}} D{  }\mathcal{A} \Omega _{D-2} Q^{-1}{dQ}\right)}{{  }\left(2 (D-1) \left(1+f'\left(R_0\right)\right) r_h{}^{2+D}+l^2\left(2 (D-3) \left(1+f'\left(R_0\right)\right)r_h{}^D+2^{\frac{D}{4}} r_h{}^2 \mathcal{A}\right)\right) \Omega _{D-2}}, \label{eq4.7}
\end{align}
where
\begin{equation}
\mathcal{B}=\pi ^{\frac{D}{D-2}} \left(-\frac{(-1)^{-\frac{D}{4}} 2^{5-\frac{D}{4}} \sqrt{1+f'\left(R_0\right)} Q}{D{  }\Omega _{D-2}}\right){}^{\frac{2}{D-2}}.\label{eq4.8}
\end{equation}
Substituting equation (\ref{eq4.7}) into equation (\ref{eq4.3}), which yields  $dS_h$ is
\begin{align}
{dS}_h=\frac{\left(1+f'\left(R_0\right)\right) l^2 r^D \left(32 \pi  p^r_h Q r+32 \sqrt{1+f'\left(R_0\right)} Q \mathcal{B} {dQ}+2^{\frac{D}{4}} D{  }\mathcal{A} \Omega _{D-2}{dQ}\right)}{4 Q \left(2 (D-1) \left(1+f'\left(R_0\right)\right) r^{2+D}+l^2 \left(2 (D-3) \left(1+f'\left(R_0\right)\right) r^D+2^{\frac{D}{4}} r^2 \mathcal{A}\right)\right)}.  \label{eq4.9}
\end{align}
In addition, we chose $D=4p$, i.e, $D = 4, 8, 12,...$, as we mentioned already.  Therefore, we will consider $D=4p$ in what follows.

Incorporating equations (\ref{eq2.7}) and (\ref{eq4.9}),   we get
\begin{align}
 T_h {dS_h}=p^r_h.  \label{eq4.10}
\end{align}
We further discuss the  thermodynamics of black hole. From equations (\ref{eq2.7}), (\ref{eq2.9}) and (\ref{eq4.9}), we have
\begin{align}
dM=\Phi_h dQ + T_h dS_h.  \label{eq4.11}
\end{align}
Obviously, one can see that as the  charged particle dropped into a  higher dimensional charged $f(R)$ black hole, the first law of thermodynamics is valid in the normal phase space. That is, equation (\ref{eq4.11}) has  evidenced that the coincidence between the variation of  $D$-dimensional $f(R)$ black hole and the first law of thermodynamics under the charged particle absorption.

Since the absorption is an irreversible process, the entropy of  final state  should be greater than  initial state of the black hole. In other words, the variation of  entropy should  satisfy  $dS_h > 0$ under the charged particle absorption. Therefore, we will check the validity of the second law of thermodynamics by equation (\ref{eq4.9}).

For the extremal black holes, we find the variation of the entropy is divergent. The divergence of the variation of entropy is meaningless. Therefore, we mainly focus on the case of near-extremal black holes, and   study the variation of entropy  numerically in the restrictions which are $D = 4p$ and $1 + f'(R_0) > 0$.  It is worth noting that the meaningful critical specific volume  exist only when $p$ is odd, it means $ p = 4, 8, 12,...$
Here, we set $Q = 1.5$ and $\Omega _{d-2}=l = p^r_h = 1$ as example. When $f'\left(R_0\right)=-0.8$, $f'\left(R_0\right)=-0.5$ and $f'\left(R_0\right)=0.5$ respectively, we get the corresponding extremal mass for  different values of $p$.  The mass of the non-extremal black hole should be larger than that of the extremal black hole. Hence, the corresponding  values of $r_h$ and $dS_h$  for different values of mass $M$ are obtained too, as shown in Table 1, Table 2, and Table 3.
\begin{center}
{\footnotesize{\bf Table 1.} The relation between ${dS}_h$, $M$ and $r_h$ of $p=1$.\\
\vspace{1mm}
\begin{tabular}{ccc|ccc|ccc}
\hline
 $f'\left(R_0\right)=-0.8$     &             &     &  $f'\left(R_0\right)=-0.5$ & & &  $f'\left(R_0\right)=0.5$ &  \\ \hline
$M$       &$r_h $  & $dS_h $&  $M$ &$r_h $& $dS_h $  &  $M$ &$r_h $& $dS_h $  \\
\hline
1.168928     &3.27626 & $410.57$       & 2.922318 &3.27488 &908.91      &8.766953    &3.27442&1527.9  \\
1.169        & 3.29521    & $48.5158 $     & 2.923&3.31530&25.2907        & 8.767    &3.28007&163.50   \\
1.17         & 3.35631     & $12.9644$     & 2.93 & 3.41385 &7.83383        & 8.77   &3.32447&20.805  \\
1.2          & 3.72504 & $2.75001 $           & 2.95 &3.54121 &4.32582         &8.78 &3.37899&10.273 \\
1.3          & 4.21406    & $1.54538 $       & 2.98&3.66171& 3.1249           &8.8   &3.44173&6.6116 \\
1.5          &4.78312    & $1.11093$       & 3.0 &3.72504 & 2.75001           &8.9   &3.61329&3.5054\\
1.8          &5.36741 &$0.89819$           & 4.0 & 5.00001& 1.01689           &9.0   &3.72504&2.7500 \\
\hline
\end{tabular}}
\end{center}
\begin{center}
{\footnotesize{\bf Table 2.} The relation between ${dS}_h$, $M$ and $r_h$ of $p=3$.\\
\vspace{1mm}
\begin{tabular}{ccc|ccc|ccc}
\hline
 $f'\left(R_0\right)=-0.8$     &             &     &  $f'\left(R_0\right)=-0.5$ & & &  $f'\left(R_0\right)=0.5$ &  \\ \hline
$M$       &$r_h $  & $dS_h $&  $M$ &$r_h $& $dS_h $  &  $M$ &$r_h $& $dS_h $  \\
\hline
0.595532   & 0.96297   & $112.422$    &1.067116  &0.93129&238.571 &2.147727  &0.89475&439.291\\
0.596       & 0.97437   & $4.95563 $   &1.068    &0.94289&4.85363 &2.148     &0.89913&11.7763 \\
0.65        & 1.07592    & $0.87591$   &1.07     &0.95219&2.87462  &2.15     &0.90747&4.35482 \\
0.68        &1.0994      & $0.80227 $  &1.25     &1.07475&0.78099   &2.18    &0.94108&1.47651 \\
0.7         & 1.11213    & $0.77289$   &1.5      &1.1338 &0.69313   &2.2     &0.95287&1.25892  \\
0.8         &1.15853     & $0.70133$   &1.8      &1.17755&0.65818    &2.5    &1.03055&0.79185 \\
0.9         &1.19062     & $0.67074$    &2.0     &1.19955&0.64537    &2.8    &1.0692 &0.72313  \\
\hline
\end{tabular}}
\end{center}
\begin{center}
{\footnotesize{\bf Table 3.} The relation between ${dS}_h$, $M$ and $r_h$ of $p=5$.\\
\vspace{1mm}
\begin{tabular}{ccc|ccc|ccc}
\hline
 $f'\left(R_0\right)=-0.8$     &             &     &  $f'\left(R_0\right)=-0.5$ & & &  $f'\left(R_0\right)=0.5$ &  \\ \hline
$M$       &$r_h $  & $dS_h $&  $M$ &$r_h $& $dS_h $  &  $M$ &$r_h $& $dS_h $  \\
\hline
0.550584  & 0.94753  & $0.71137$   &0.935858    &0.92729  &0.71331  &1.7437     &0.87262&34.5235  \\
0.551     & 0.94834  & $0.69971 $  &0.936       &0.92745  &0.71090  &1.768      &0.90363&0.71417   \\
0.56      & 0.96185  & $0.56598$   &0.94        &0.93164  &0.65539  &1.77       &0.90479&0.69672   \\
0.6       & 0.99315  & $0.44083 $  &0.95        &0.94002  &0.57657  &1.8        &0.91802&0.56245  \\
0.7       & 1.02988  & $0.38878 $  &0.98        &0.95705  &0.48544  &1.9        &0.94275&0.45323   \\
0.8       & 1.05127  & $0.37393$   &0.99        &0.96136  &0.47044  &2.0        &0.95800&0.42044   \\
0.9       & 1.06674  & $0.36657$   &2.0         &1.07275  &0.35823  &3.0        &1.02204&0.36658   \\
\hline
\end{tabular}}
\end{center}

From these tables, the value of  extremal mass  changed  with the value  of  $f'(R_0)$. It can be seen that as the value of $f'(R_0)$ increases, the extremal mass and $dS_h$  increases too, but the value of $r_h$  decreases. In addition, the event horizon of the black hole increases  with the variation of mass for the same $f'(R_0)$, but the value of $dS_h$ decreases.  Fortunately, the results show that when the mass is greater than the extremal mass, the variation of  entropy is always positive . This implies that the second law of thermodynamics is valid for the  near-extremal $f(R)$ black holes  in the normal phase space.

In order to make our result clearer and more visible, we also can obtain the relation between $dS_h$ and $r_h$ under the condition $-1<f'\left(R_0\right)$ for different values of $p$, which  shown in Figure 1.
It  shows clearly that the value of $dS_h$ decreases  with the variation of $r_h$, but there is always $dS_h>0$. This result is consistent with the result of the above tables.  So, the second law of thermodynamics is valid for  $D$-dimensional $f(R)$ charged black hole (near- or non-extremal case ) in the normal phase space.
\begin{figure}[htb] \centering\subfigure[{$p=1$.}] {\includegraphics[scale=0.55,keepaspectratio]{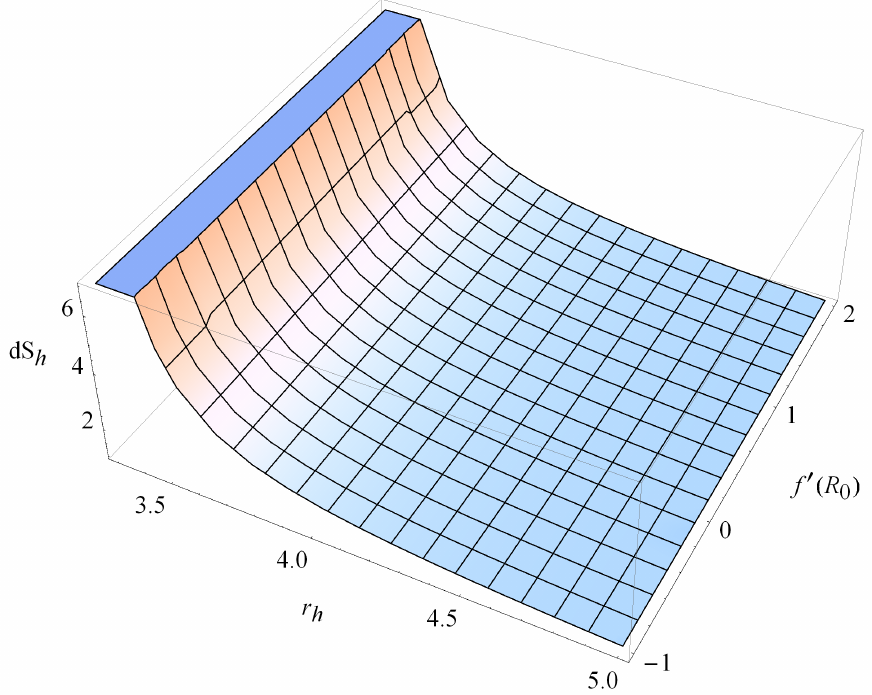}} \quad \centering\subfigure[{$p=3$.}] {\includegraphics[scale=0.55,keepaspectratio]{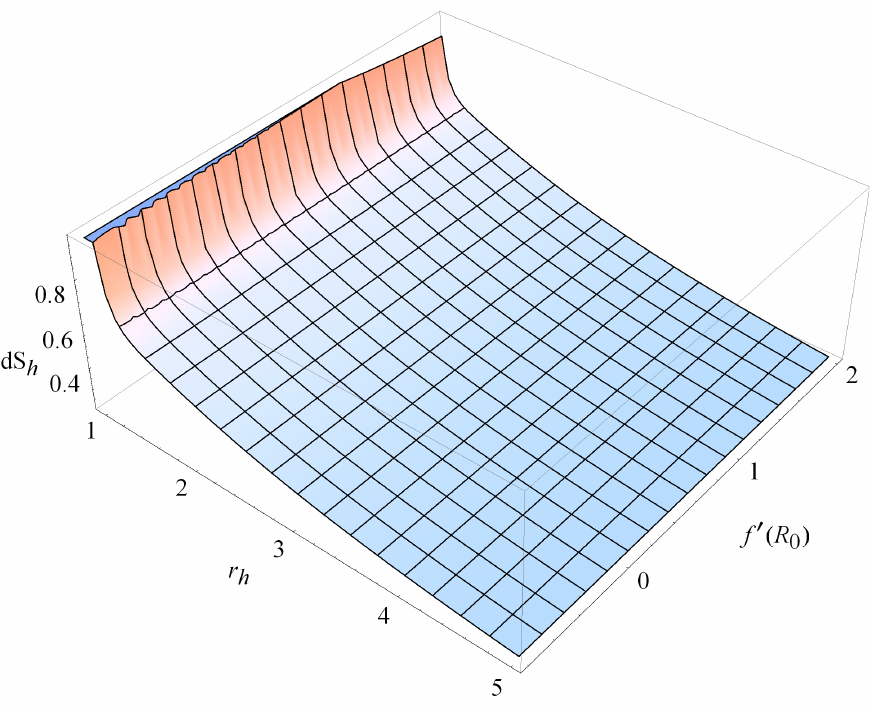}} \quad \centering\subfigure[{$p=5$.}] {\includegraphics[scale=0.60,keepaspectratio]{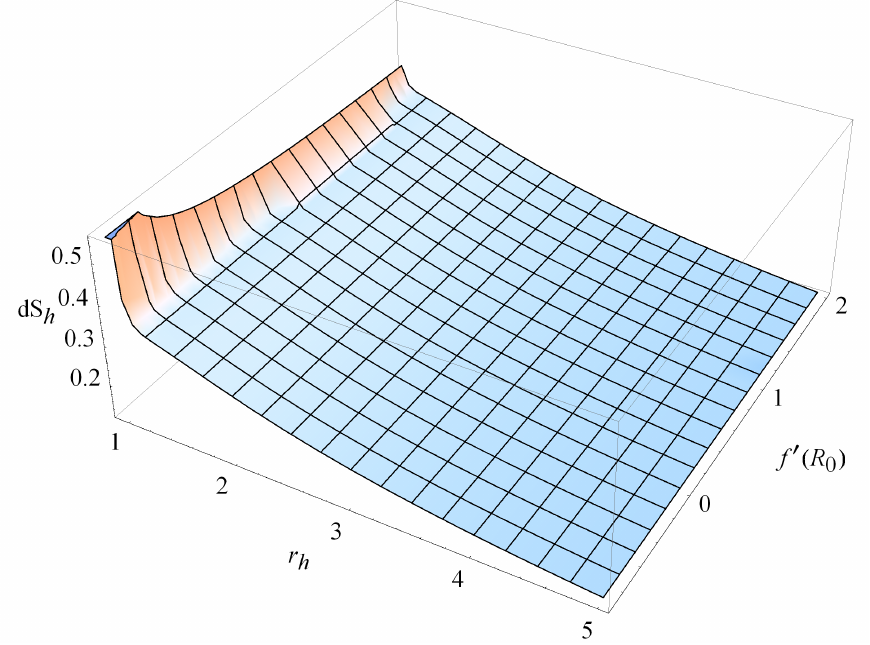}}\\
\caption{The relation between $dS_h$ and $r_h$ for the condition $-1<f'\left(R_0\right)$.}
\label{fig:1}
\end{figure}

\subsection{Thermodynamics in the extended phase space}
In the extended  phase space, since we are going to discuss the thermodynamics of the black hole  by introducing the pressure, the mass should be interpreted as enthalpy rather than internal energy. In addition, other thermodynamic quantities can be obtained through thermodynamic identities. In other words, we should use equation (\ref{eq2.15}), that is
\begin{align}
M=U_h+PV_h.  \label{eq4.13}
\end{align}
Therefore, based on the energy conservation and charge conservation, the energy relation of equation (\ref{eq3.9}) becomes
\begin{align}
\omega=dU_h=d(M-PV_h), \quad e=dQ \label{eq4.14}
\end{align}
Then, we can obtain
\begin{align}
d(M-{PV_h})=\frac{q}{r_h}\sqrt{1+f'\left(R_0\right)}{dQ}+p^r_h.  \label{eq4.15}
\end{align}
Equation (\ref{eq4.15}) is different from that in the normal phase space where the increase of energy is related to the  mass of the black hole.
Similarly, the event horizon and  function $W(h)$ will change due to the charged particle absorption. The variation of horizon radius can be obtained from the variation of metric function $W (M, Q, l, r_h)$. So, the $dW_h$ is,
\begin{align}
dW_h=\frac{\partial W_h}{\partial M}{dM}+\frac{\partial W_h}{\partial Q}{dQ}+\frac{\partial W_h}{\partial r_h}{dr}_h+\frac{\partial W_h}{\partial l}{dl}=0,\quad {W_h}=f(M,Q,l,r_h), \label{eq4.16}
\end{align}
and
\begin{align}
&\frac{\partial W_h}{\partial M}=-\frac{16\text{  }\pi  {r_h}^{3-D}}{(D-2) \left(1+f'\left(R_0\right)\right)\Omega _{D-2}}{dM},  \quad \nonumber\\
&\frac{\partial W_h}{\partial Q}=-\frac{2^{\frac{D}{4}-1} D{  }r_h{}^{2-D}{  }\mathcal{A}}{(D-2) \left(1+f'\left(R_0\right)\right) Q}{dQ},\quad \nonumber\\
&\frac{\partial W_h}{\partial r_h}=\frac{1}{2} r_h \left(\frac{4}{l^2}+\frac{r_h{}^{-D} \left(32 (D-3) M \pi  r_h+2^{\frac{d}{4}} (D-2)^2 \mathcal{A} \Omega _{D-2}\right)}{(D-2) \left(1+f'\left(R_0\right)\right) \Omega _{D-2}}\right){dr}_h, \quad \nonumber\\
&\frac{\partial W_h}{\partial l}=-\frac{2r_h{}^2}{l^3}{dl}.  \label{eq4.17}
\end{align}
Substituting equation (\ref{eq4.15}) into equation (\ref{eq4.16}), we get
\begin{align}
&{dr}_h=\frac{16 (-1)^{-\frac{D}{4}} r_h{}^{2-D}}{(D-1) \left(1+f'\left(R_0\right)\right)\text{  }\Omega _{D-2} } \quad \nonumber\\
&~~~~~\times \frac{ \left(p^r_h r_h \mathcal{A} \mathcal{C}-r_h \left(-(-1)^{\frac{D}{4}} \pi +\mathcal{A} \mathcal{C}\right){dM}+{  }X\right)}{ \left(\frac{D-3}{ r_h}+(-1)^{-\frac{D}{4}} (D-1){  }r_h \left((-1)^{\frac{D}{4}}- \mathcal{A} \mathcal{D}\right)l^{-2}+\left(1+f'\left(R_0\right)\right){}^{-1}2^{\frac{D}{4}-1}{  }r_h{}^{1-D} \mathcal{A}\right)}, \label{eq4.18}
\end{align}
where
\begin{align}
\mathcal{C}=\pi ^{-\frac{2}{-2+D}} \left(-\frac{(-1)^{-D/4} 2^{5-\frac{D}{4}} \sqrt{1+f'\left(R_0\right)} Q}{D \Omega _{D-2}}\right){}^{\frac{D}{2-D}},\label{eq4.19}
\end{align}
\begin{align}
\mathcal{D}=\pi ^{\frac{D}{2-D}}\left(-\frac{(-1)^{-D/4} 2^{5-\frac{D}{4}} \sqrt{1+f'\left(R_0\right)} Q}{D \Omega _{D-2}}\right){}^{\frac{D}{2-D}},\label{eq4.20}
\end{align}
\begin{align}
X= (2l)^{-3}(D-2)\text{  }\left(1+f'\left(R_0\right)\right) r^D \left((-1)^{\frac{D}{4}}- \mathcal{A} \mathcal{D}\right)\Omega _{D-2}{dl}.\label{eq4.21}
\end{align}
From equation (\ref{eq4.18}),  the variations of  entropy and volume can be expressed as
\begin{align}
{dS}_h=\frac{4r_h{}^{-1}(-1)^{-\frac{D}{4}} \left(\text{   }p^r_h r_h \mathcal{A} \mathcal{C}- r_h \left(\mathcal{A} \mathcal{C}-(-1)^{\frac{D}{4}} \pi \right){dM}+X \right)}{ \frac{D-3}{r_h}+(-1)^{-\frac{D}{4}} (D-1) r_h \left((-1)^{\frac{D}{4}}- \mathcal{A} \mathcal{D}\right)l^{-2}+\left(1+f'\left(R_0\right)\right){}^{-1}2^{\frac{D}{4}-1}r_h{}^{1-D} \mathcal{A}},  \label{eq4.22}
\end{align}
and
\begin{align}
{dV}_h=\frac{16 (D-2)^{-1}(-1)^{-\frac{D}{4}} \left(\left|p^r_h\right| r_h \mathcal{A} \mathcal{C}-r \left( \mathcal{A} \mathcal{C}-(-1)^{\frac{D}{4}} \pi \right){dM}+X\right)}{\frac{D-3}{r_h}+(-1)^{-\frac{D}{4}} (D-1) r_h \left((-1)^{\frac{D}{4}}- \mathcal{A} \mathcal{D}\right)l^{-2}+\left(1+f^,\left(R_0\right)\right){}^{-1}2^{-1+\frac{D}{4}} r_h{}^{1-D}\mathcal{A}}.\label{eq4.23}
\end{align}
With the help of  equations  (\ref{eq2.7}), (\ref{eq2.11}), (\ref{eq4.22}) and (\ref{eq4.23}), we can get the following relation
\begin{align}
{T_hdS}_h-{PdV}_h=p^r_h,  \label{eq4.24}
\end{align}
and
\begin{align}
dU=\Phi dQ+T_h {dS}_h-{P dV}. \label{eq4.25}
\end{align}
We can prove that the above physical quantities satisfy the first law of thermodynamics. That is,
\begin{align}
{dM}=T_h{dS}_h+\Phi _h{dQ}+V_h{dP},  \label{eq4.26}
\end{align}
which is consistent with equation(\ref{eq2.13}). Therefore, the first law of thermodynamics in the higher-dimensional charged $f(R)$ black hole is well recovered when a charged particle is absorbed.

The satisfaction of the first law of thermodynamics does not mean that the second law is also satisfied, especially in the extended phase space. Hence, we will investigate  the  second law of thermodynamics of the  higher-dimensional charged $f(R)$ black hole by use  equation (\ref{eq4.22}) in the extended phase space. For the extremal black holes, the variation of  entropy takes on the form
\begin{equation}
{dS}_h=-\frac{4\pi p^r_h l^2}{(D-1)r_h}. \label{eq4.27}
\end{equation}
In  equation (\ref{eq4.27}), there is a minus sign. That is, the entropy decreases in the chronological direction for the extremal black hole, and this result does not support  the second law of thermodynamics under the consideration of $PV_h$ term.
Then, we focus on the near-extremal black holes. Similarly, since equation (\ref{eq4.22}) includes a bunch of parameters, for simplicity and
without loss of generality, we also set $l=p^r_h=\Omega _{d-2}=1$ and $Q=1.5$. For different values of the parameters $p$ and $f'(R_0)$, we can get different mass of the extremal black holes, and  the corresponding value of  $r_h$ and $dS_h$ are also obtained, which are listed in Table 4, Table 5, and Table 6.
\begin{center}
{\footnotesize{\bf Table 4.} The relation between $\text{dS}_h$, $M$ and $r_h$ of $p=1$.\\
\vspace{2mm}
\begin{tabular}{ccc|ccc|ccc}
\hline
 $f'\left(R_0\right)=-0.8$     &             &     &  $f'\left(R_0\right)=-0.5$ & & &  $f'\left(R_0\right)=0.5$ &  \\ \hline
$M$       &$r_h $  & $dS_h $&  $M$ &$r_h $& $dS_h $  &  $M$ &$r_h $& $dS_h $  \\
\hline
1.168927    & 3.27393 & $-1.2797$ & 2.922318    &3.27488 &-1.2809 &8.766953    &3.27442 &-1.2803  \\
1.17        & 3.35631 & $-1.3809$ & 3.5         &4.53012 &-3.4896 &8.767       &3.28007 &-1.2871   \\
5.5         & 8.62368 & $-28.687$ & 10.5        &7.79934 &-20.246 &10.5        &4.53012 &-3.4896  \\
53.25        & 18.8087 & $-54346$ & 130        &18.6595 &-11452 &395.5       &18.7473 &-21537 \\
55.5        & 19.0714 & $10361.7$ & 135       &18.8971 &47252 &405.5       &18.9049 &40634 \\
65.5        & 20.1597 & $2014.54$ & 155       &19.8173 &2613.63 &415.5       &19.0599 &10911\\
75.5        & 21.1416 & $1295.5$  & 200         &1151.23 &21.5552 &515.5       &20.4876 &1677 \\
\hline
\end{tabular}}
\end{center}
\begin{center}
{\footnotesize{\bf Table 5.} The relation between $\text{dS}_h$, $M$ and $r_h$ of $p=3$.\\
\vspace{2mm}
\begin{tabular}{ccc|ccc|ccc}
\hline
 $f'\left(R_0\right)=-0.8$     &             &     &  $f'\left(R_0\right)=-0.5$ & & &  $f'\left(R_0\right)=0.5$ &  \\ \hline
$M$       &$r_h $  & $dS_h $&  $M$ &$r_h $& $dS_h $  &  $M$ &$r_h $& $dS_h $  \\
\hline
0.595532    & 0.96297   & $-1.1990 $   &1.067116&0.93129&-1.2330   &2.147727&0.89475&-1.2805\\
0.599       & 0.99412   & $-2.5564 $   &1.068&0.94289&-1.6147      &2.148&0.89913&-1.4242  \\
0.6         & 0.99823   & $-2.9$        &1.1&0.99852&-59.821       &2.2015&0.95364&-29.968 \\
0.6075      & 1.01954   & $-7.1647 $   &1.10225&1.00061&-655.3     &2.20955&0.95758&-595.4 \\
0.6175      & 1.03807   & $-1302.7$    &1.1025&1.00083&20355       &2.210985 &0.95826&276.22 \\
0.617925    & 1.03873   & $277.301$     &1.2&1.05667&3.52123       &2.5&1.03055& 2.77183\\
0.8         &1.15853    & $2.42865$       &1.5&1.1338&2.2210       &2.8&1.0692& 2.23741\\
\hline
\end{tabular}}
\end{center}
\begin{center}
{\footnotesize{\bf Table 6.} The relation between $\text{dS}_h$, $M$ and $r_h$ of $p=5$.\\
\vspace{2mm}
\begin{tabular}{ccc|ccc|ccc}
\hline
 $f'\left(R_0\right)=-0.8$     &             &     &  $f'\left(R_0\right)=-0.5$ & & &  $f'\left(R_0\right)=0.5$ &  \\ \hline
$M$       &$r_h $  & $dS_h $&  $M$ &$r_h $& $dS_h $  &  $M$ &$r_h $& $dS_h $  \\
\hline
0.550584   &0.94753   & $-37.169$  &0.935858  &0.92729&-7708 &1.743696  &0.87231&-0.7639\\
0.550985   &0.94831   & $-181.37$  &0.93589   &0.92732&1389.57 &1.744     &0.87596&-0.9112\\
0.55122    &0.94875   & $156.465$  &0.9359    &0.92734&777.866 &1.76589   &0.90235&-259.54\\
0.55211    &0.95038   & $20.2037$  &0.94      &0.93164&8.53264 &1.768589  &0.90398&22.67  \\
0.56       &0.96185   & $3.19932$  &1.0       &0.96526&1.38722 &1.8       &0.91802&2.56467\\
0.6        &0.99315   & $1.30405$  &1.25      &1.01645&0.92943 &2         &0.95800&1.07527\\
0.985      &1.07729   & $0.88591$  &1.5       &1.04181&0.87806 &5         &1.07024&0.82968\\
\hline
\end{tabular}}
\end{center}
From these tables, we find that the variation of entropy is more sophisticated, and the value of $dS_h$ is  not a simple monotonic relationship such as that in  the normal phase space, there is always a divergent point. Although the value of $dS_h$  decrease as the mass increase, the value of $dS_h$ have positive and negative regions.  When the mass approaches to extremal mass, the value of $dS_h$ is negative, which means that the second law of thermodynamics is invalid for the near-extremal $f(R)$ black hole. In the other hands, when the mass is larger than the extremal mass, the change of entropy is positive, therefore, this result supports the second law of thermodynamics for the non-extremal $f(R)$ black hole. In addition, the result  demonstrates that the location of the divergence point is different when the value of $f'(R_0)$ is changed, that is, there is a great connection between the divergence point and the value of $f'(R_0)$. When the value of $f'(R_0)$  increased, the later the divergent point appears. The relation between $dS_h$, $r_h$ and $f'(R_0)$ can be plotted while $p$ is change, which is shown in Figure 2.
\begin{figure}[htb] \centering\subfigure[{$p=1$.}] {\includegraphics[scale=0.55,keepaspectratio]{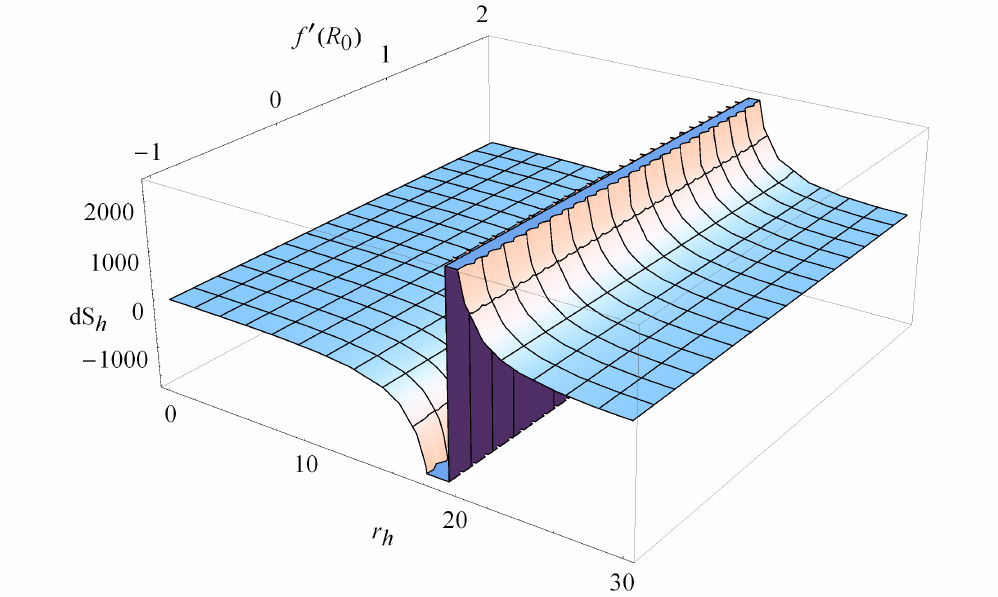}} \quad \centering\subfigure[{$p=3$.}] {\includegraphics[scale=0.50,keepaspectratio]{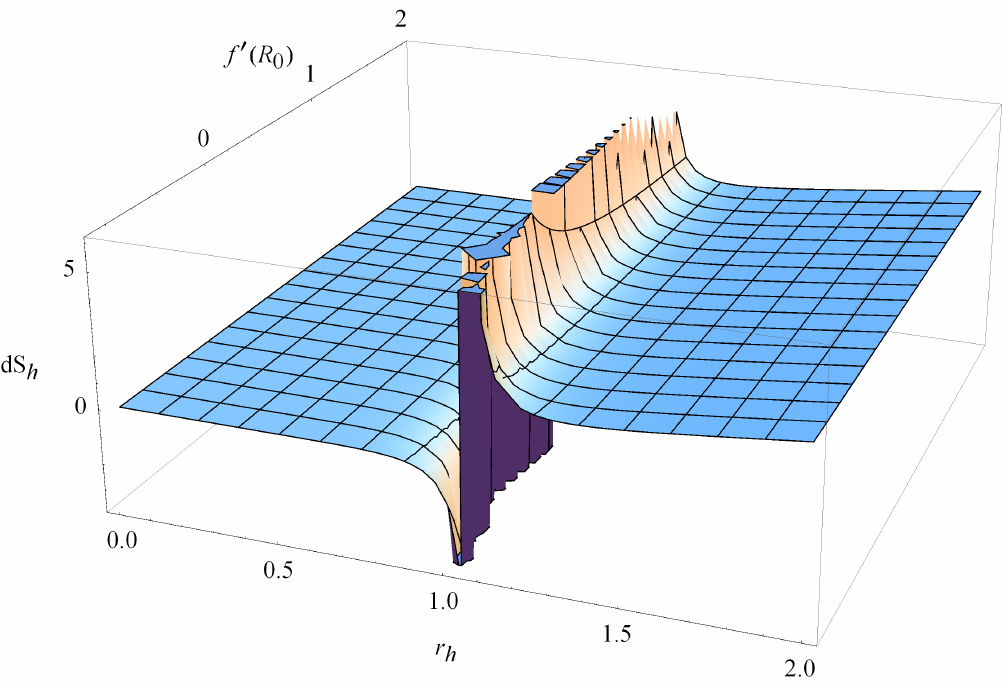}} \quad \centering\subfigure[{$p=5$.}] {\includegraphics[scale=0.50,keepaspectratio]{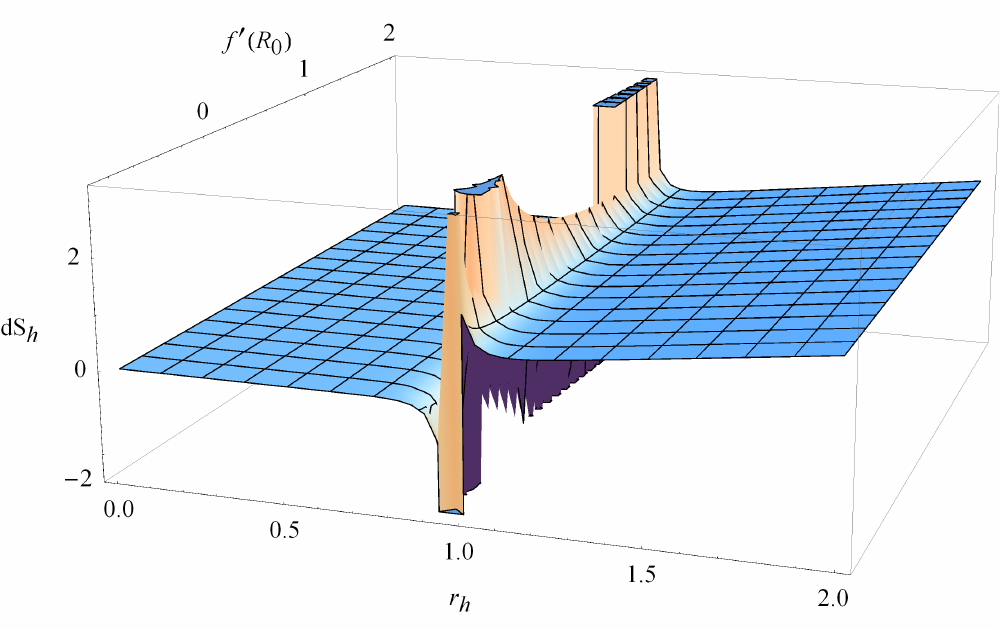}}\\
\caption{The relation between $dS_h$ and $r_h$ for the condition $-1<f'\left(R_0\right)$.}
\label{fig:2}
\end{figure}

From these figures, we find that there  is always a phase transition point  which divides $dS_h$ into positive and negative  regions, this result is consistent with the conclusion of the above tables. Furthermore, the result shows that $dS_h$ is negative when the event horizon radius is smaller than the phase transition point. Therefore, we  can also  conclude that  the second law of thermodynamics  is not valid in the extended phase space for the near-extremal black hole  under charged particle absorption. Obviously, the result also shows that the magnitudes of the violation for the second law of thermodynamics is related to  the parameters $l, p, Q, f'(R_0), \Omega _{d-2}$.

\section{The weak cosmic censorship conjecture of the higher dimensional  $f(R)$ black holes }
\label{sec:5}
In this section, we investigate the validity of the weak cosmic censorship conjecture for the higher-dimensional  $f(R)$ black hole, and we intend to explore  what the final state is as the charged particle is absorbed by the higher-dimensional charged $f(R)$ black hole in  different  phase spaces. As the extremal black hole is in a state in which its  mass has the maximum charge, it is feasible to overcharge the black hole by adding the charged particle. In the other words, the event horizon will disappear, which makes the singularity of the black hole  exposed in the spacetime. Hence, we should check whether there is an event horizon  at the final state of the black hole. For the black hole, the metric function $W(r)$  has a  minimum point  $W(r_{\min })$.  And, there at least is a positive real root for the equation  $W(r_{\min })=0$, the final states still black hole and the weak cosmic censorship conjecture still holds. Otherwise, the weak cosmic censorship conjecture is invalid. So, near the locations of the minimum value $r_{\min }$, the following relations are satisfied
\begin{align}
W(r)|_{r=r_\text{min}}\equiv W_\text{min}=\delta\leq 0,\quad \partial_{r}W(r)|_{r=r_\text{min}}\equiv W'_\text{min}=0, \quad
(\partial_{r})^2 f(r)|_{r=r_\text{min}}>0.  \label{eq5.1}
\end{align}
The minimum value of the function $W(r)$ is $\delta$. For the case of extremal black hole $\delta=0$,  and the location of the event horizon is coincident with that of the minimum value of the function $W(r_{\min })$. For the case of  near-extremal black hole, $\delta$ is a very small negative value. When  charged particle dropped  into the black hole, the change of the conserved quantities of the black hole  can be written as $W(M+dM, Q+dQ, l+dl)$. Correspondingly, the position of the minimum point of function $W(r_{\min })$ and  event horizon change into $r_{\min}\rightarrow r_{\min}+dr_{\min}, r_h\rightarrow r_h+dr_h$ respectively. Then, there is also a shift for the value of $W(r_{\min })$, which is denoted  as $dW_{\min}$. At the new lowest point , we have
\begin{align}
\partial _rW|_{r=r_{\min }+{dr}_{\min }}=W'_{\min }+{dW'}_{\min }=0.  \label{eq5.2}
\end{align}

\subsection{  Weak cosmic censorship conjecture in the normal phase space }
In the normal phase space,  we will study the change of $W(r_\text{min })$ as  charged particle  absorbed. At $r_\text{min }+{dr}_\text{min }$,  with the help of condition $W'_{\min }=0$ in equation (\ref{eq5.2}), we have a relation ${dW'}_{\min }=0$, which implying
\begin{align}
{dW}'_\text{min }=\frac{\partial W'_\text{min }}{\partial M} dM+\frac{\partial W'_\text{min }}{\partial Q} dQ+\frac{\partial W'_\text{min }}{\partial r_\text{min }} dr_\text{min }=0.  \label{eq5.3}
\end{align}
In addition, at the new minimum point, $ W\left(r_\text{min }+dr_\text{min }\right)$ can be expressed as
\begin{align}
W\left(r_\text{min }+dr_\text{min }\right)=W_\text{min }+dW_\text{min }, \label{eq5.4}
\end{align}
where
\begin{align}
dW_\text{min }=\frac{\partial W_\text{min }}{\partial M} dM+\frac{\partial W_\text{min }}{\partial Q} dQ. \label{eq5.5}
\end{align}
For the extremal black hole, $W_\text{min }=\delta =0$ and the temperature is zero $T_h=0$.  Substituting  equation (\ref{eq4.2}) into equation (\ref{eq5.5}), we can get
\begin{equation}
{dW}_{\min }=0. \label{eq5.6}
\end{equation}
This implies that  $W_\text{min }+dW_\text{min }=0$, which means that  the final state of the extremal black hole still an extremal black hole with the new mass and charge when  particle  absorbed.  Hence, the existence of the event horizon ensures that the singularity is not naked in this black hole, and the weak cosmic censorship conjecture is valid.
For the near-extremal black hole, we have
\begin{align}
r_h=r_{\min}+\epsilon, \quad  \delta\rightarrow\delta _{\epsilon },  \label{eq5.7}
\end{align}
where $0<\epsilon\ll1$, and the minimum value $\delta _{\epsilon }$ is a very small negative value with respect to $\epsilon$.
Then,  the equation (\ref{eq4.2}) is rewritten in terms of $\epsilon$ and $r_{\min }$, which is
\begin{align}
&{dM}=\frac{\mathcal{B}\text{  }\sqrt{1+f'\left(R_0\right)}\text{  }{dQ}}{ \pi  r_{\min }}+\frac{(D-1) \left(1+f'\left(R_0\right)\right) r_{\min }{}^{2+D}\Omega _{D-2}}{16l^2 \pi  r_{\min }{}^4(D-2)^{-1}}{dr}_{\min } \quad \nonumber\\
&~~~~~+\frac{ \left(2^{\frac{D}{4}} \mathcal{A} r_{\min }{}^2+2 (D-3) \left(1+f'\left(R_0\right)\right) r_{\min }{}^D\right) \Omega _{D-2}}{32\text{  }\pi  r_{\min }{}^4(D-2)^{-1}}{dr}_{\min } \quad \nonumber\\
&~~~~~-\frac{\mathcal{B}\text{  }\sqrt{1+f'\left(R_0\right)}\text{  }\epsilon }{ \pi  r_{\min }{}^2}{dQ}+\frac{ (D-1) \left(1+f'\left(R_0\right)\right) r_{\min }{}^{2+D} \Omega _{D-2}\epsilon  }{16 l^2 \pi  r_{\min }{}^5(D-2)^{-1}}{dr}_{\min } \quad \nonumber\\
&~~~~~-\frac{\left(2^{\frac{D}{4}} \mathcal{A}(D-1) r_{\min }{}^2+2 (D-3) \left(1+f'\left(R_0\right)\right) r_{\min }{}^D\right) \Omega _{D-2}\epsilon  }{32\text{  }\pi  r_{\min }{}^5(D-2)^{-1}}{dr}_{\min }+\mathcal{O}(\epsilon)^2. \label{eq5.8}
\end{align}
Substituting equation (\ref{eq5.8}) into equation (\ref{eq5.5}), and consider the condition $D = 4p, p\in \mathbb{N}$, we can obtain
\begin{equation}
{dW}_{\min }= \mathcal{O}(\epsilon)^2. \label{eq5.9}
\end{equation}
Then, for the near-extremal black hole, the  equation (\ref{eq5.4}) becomes
\begin{align}
W\left(r_\text{min }+dr_\text{min }\right)=\delta _{\epsilon }+\mathcal{O}(\epsilon)^2. \label{eq5.10}
\end{align}
For the  special case  where $\epsilon=0$ in  equation(\ref{eq5.10}), we can have $ W\left(r_\text{min }+dr_\text{min }\right)=0$. Interestingly, this result is consistent with the result of the extremal case. Hence,  equation (\ref{eq5.6}) is further confirmed. However, we still does not estimate the value magnitudes between $\left|\delta _{\epsilon }\right|$ and $\mathcal{O}(\epsilon)^2$, when the value of $\delta _{\epsilon }$ was not zero. Therefore, for the near-extremal black hole, we can not simply ignore the contribution of $\mathcal{O}(\epsilon)^2$ to equation (\ref{eq5.10}) since $\delta _{\epsilon }$ is also a small quantity, so we need a more precise calculation. To the second order, we find
\begin{align}
&W\left(r_{\min }+\epsilon \right)=\frac{ (D-1)\text{  }r_{\min }{}^2}{ (D-3)\text{  }l^2 }+\frac{ 2 (D-3) \left(1+f'\left(R_0\right)\right)\text{  }r_{\min }{}^D+2^{\frac{D}{4}}\text{  }r_{\min }{}^2 \mathcal{A}}{2 (D-3) \left(1+f'\left(R_0\right)\right)\text{  }r_{\min }{}^D} \quad \nonumber\\
&~~~~~~~~~~~~~~~~+\frac{ \left(4 (-1+D) \left(1+f'\left(R_0\right)\right)\text{  }r_{\min }{}^D-2^{D/4} (D-2) l^2 \mathcal{A}\right) \epsilon ^2}{4 r_{\min }{}^D \left(1+f'\left(R_0\right)\right) l^2}+\mathcal{O}(\epsilon)^3.\label{eq5.11}
\end{align}
Hence, we can get
\begin{align}
\delta_\epsilon=-\frac{ \left(4 (-1+D) \left(1+f'\left(R_0\right)\right)\text{  }r_{\min }{}^D-2^{D/4} (D-2) l^2 \mathcal{A}\right) \epsilon ^2}{4 r_{\min }{}^D \left(1+f'\left(R_0\right)\right) l^2}-\mathcal{O}(\epsilon)^3.\label{eq5.12}
\end{align}
Similarly,  to the second order,   ${dW}_{\min }$ can be expanded as
\begin{align}
&{dW}_{\min }=-\frac{\left(4-5 D+D^2\right)\epsilon ^2}{r_{\min }l^2}\text{  }{dr}_{\min }-\frac{2^{\frac{D}{4}}\epsilon ^2}{r_{\min }{}^{(D+1)}l^2} (D-2) l^2 \left(-(8 \pi )^{\frac{4}{D-2}} \lambda _1\right){}^{\frac{D}{4}} {dr}_{\min } \quad \nonumber\\
&~~~~~~~~~+\frac{ 2^{\frac{8+6 D+D^2}{4 (D-2)}} \pi ^{\frac{D}{D-2}} \epsilon ^2 r_{\min }^{-1-D} \left(\left(6-5 D+D^2\right) l^2+(D-1) D r_{\min }^2\right) \lambda _{1 {dr}_{\min }}}{(-1)^{-\frac{D}{4}}\left((D-3) l^2+(-1+D) r_{\min }^2\right)}+\mathcal{O}(\epsilon)^3,\label{eq5.13}
\end{align}
where
\begin{align}
\lambda _1=\left(\left(-2^{-\frac{12}{D-2}} \pi ^{-\frac{4}{D-2}} \left(-\frac{2^{1-\frac{D}{4}} r_{\min }{}^{D-2} \left((D-3) l^2+(D-1) r_{\min }{}^2\right)}{l^2}\right){}^{\frac{D}{4}}\right){}^{\frac{1}{4} (D-2)}\right){}^{\frac{4}{D-2}}.\label{eq5.14}
\end{align}
It is easy to find that the relation between $\delta_\epsilon$ and $\mathcal{O}(\epsilon)^2$. For simplicity, we redefine
\begin{align}
\mathcal{W}_N=\frac{\delta _{\epsilon }+\mathcal{O}(\epsilon )^2}{\epsilon ^2}.\label{eq5.15}
\end{align}
In order to make the results  gain an intuitive understanding, the result of equation (\ref{eq5.15}) is plotted, which is shown in Figure 3.
\begin{figure}[htb] \centering\subfigure[{$p=1$.}] {\includegraphics[scale=0.50,keepaspectratio]{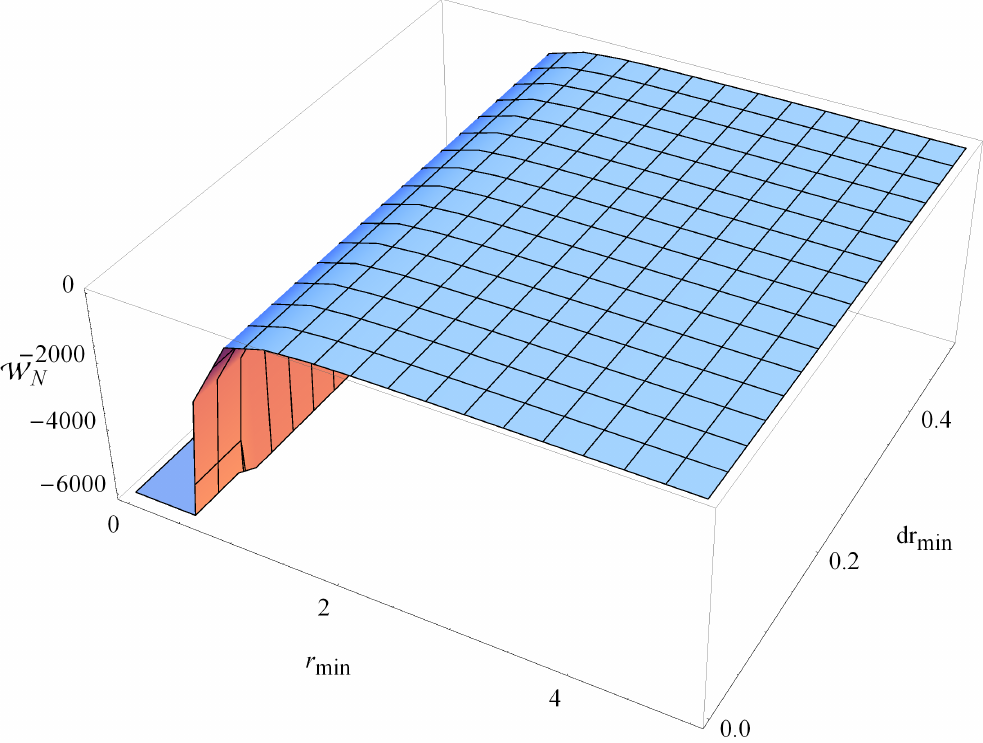}} \quad \centering\subfigure[{$p=3$.}] {\includegraphics[scale=0.58,keepaspectratio]{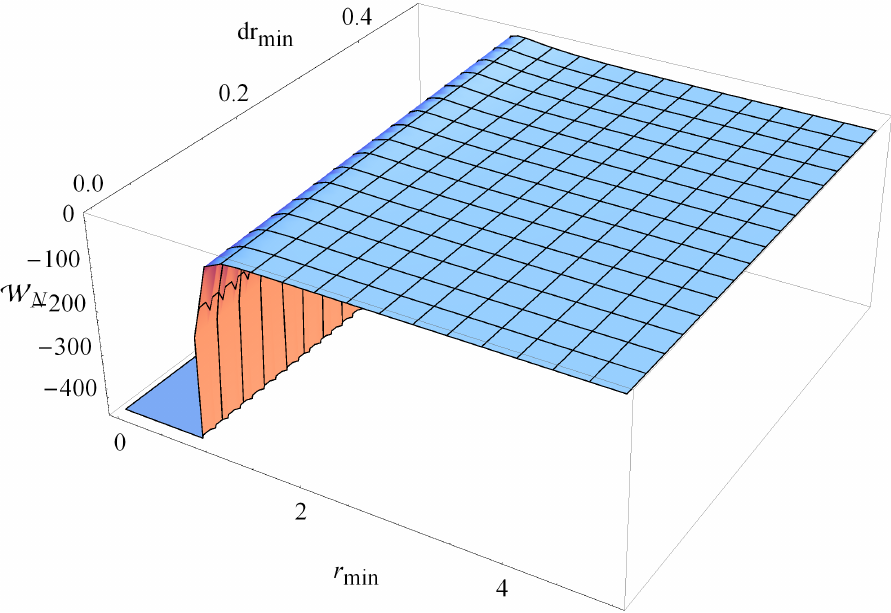}} \quad \centering\subfigure[{$p=5$.}] {\includegraphics[scale=0.58,keepaspectratio]{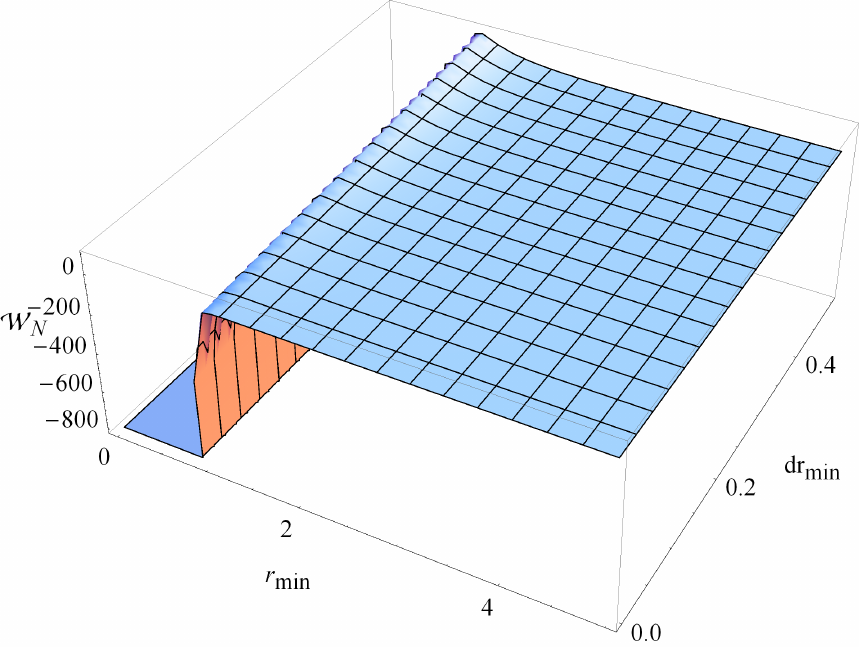}}\\
\caption{The value of $\mathcal{W}_N$ for $l = 1$, $Q = 2$, $\Omega _{D-2}$ = 1.}
\label{fig:3}
\end{figure}
 Fortunately, for different values of $p$, there is always $\mathcal{W}_N<0$ in  Figure 3. In  other words, the result shows that $W\left(r_\text{min }+dr_\text{min }\right)=\delta _{\epsilon }+\mathcal{O}(\epsilon )^2<0$, which means  the weak cosmic censorship conjecture for the near-extremal higher dimensional charged $f(R)$ black hole is valid under charged particles absorption in the normal phase space.
\subsection{ Weak cosmic censorship conjecture in the extended phase space}
In the expended phase space, $l$ is a variable which leads to  the conserved quantity such as mass $M$, charge $Q$, and AdS radius $l$ will
transform into $(M + dM, Q + dQ, l + dl)$ as a charged particle swallowed by the black hole. Therefore, according equation (\ref{eq5.2}), we  can also get
\begin{align}
{dW}'_\text{min }=\frac{\partial W'_\text{min }}{\partial M} dM+\frac{\partial W'_\text{min }}{\partial Q} dQ+\frac{\partial W'_\text{min }}{\partial r_\text{min }} dr_\text{min }+\frac{\partial W'_\text{min }}{\partial l} dl=0.  \label{eq5.16}
\end{align}
In addition, at the new minimum point, we obtain
\begin{align}
W\left(r_\text{min }+dr_\text{min }\right)=W_\text{min }+dW_\text{min }, \label{eq5.17}
\end{align}
and
\begin{align}
dW_\text{min }=\frac{\partial W_\text{min }}{\partial M} dM+\frac{\partial W_\text{min }}{\partial Q} dQ +\frac{\partial W_\text{min }}{\partial l} dl. \label{eq5.18}
\end{align}
Deserve to be mentioned, equation (\ref{eq5.17}) is dissimilar from  equation (\ref{eq5.4}) due to the emergence of the  cosmological constant. For the extremal black hole,  $r_\text{min}$   locates at $r_h$, so  equation (\ref{eq5.3}) can be applied. In this case, we also have $W_\text{min }=\delta =0$, inserting equation (\ref{eq5.3}) into equation (\ref{eq5.18}), we can  get
\begin{align}
dW_\text{min }=0. \label{eq5.19}
\end{align}
In accordance with equation (\ref{eq5.19}), we also get  $W_\text{min }+dW_\text{min }=0$. It  shows clearly that there is not any change in $W\left(r_\text{min }+dr_\text{min }\right)$ for the extremal black holes so that the black hole has  horizon after the absorption in the extended phase space. Therefore,  the weak cosmic censorship conjecture is valid  for the extremal higher-dimensional charged $f(R)$ black holes. It is interesting to note that this conclusion has not different with that in the normal phase space, the black hole keeps its configuration after the absorption. Hence, the  extremal black hole  still  extremal black hole with the contribution of pressure, that is,  the particle with sufficient momentum and charge would not overcharge extremal higher-dimensional $f(R)$ black hole in the extended phase space.

Similarly, for the near-extremal black hole, we also utilize $r_h=r_\text{min }+\epsilon $, we can expand equation (\ref{eq5.3}) at $r_\text{min }$, which leads to
\begin{align}
&dM=\frac{\text{    }\sqrt{1+f'\left(R_0\right)}{  }\pi ^{\frac{1}{8 D-1}} \lambda_2 }{ r_{\min }}{dQ}-\frac{(8 D-1)\Omega _{D-2}{   }\left(1+f'\left(R_0\right)\right) r_{\min }{}^{16 D-1}{dl}}{4 l^3 \pi  }  \quad \nonumber\\
&+\frac{(8 D-1){  }(16 D-1) \left(1+f'\left(R_0\right)\right)\Omega _{D-2}r_{\min }{}^{16 D-2}}{8l^2 \pi  }{dr}_{\min }-\frac{\text{  }\sqrt{1+f'\left(R_0\right)}\text{  }\pi ^{\frac{1}{8 D-1}}\text{  }\lambda_2}{8\text{  }r_{\min }{}^2}{dQ} \quad \nonumber\\
&+\frac{(8D-1)\left(2 (16 D-3) \left(1+f'\left(R_0\right)\right) {r_{\min }}^{16 D-2}+16^D\text{  }\left(-\pi ^{\frac{2}{8 D-1}}\lambda_2\right)^{4 D}\right)\Omega _{D-2}}{16\text{  }\pi  r_{\min }{}^2}{dr}_{\min } \quad \nonumber\\
&-\frac{(8 D-1) (16 D-1) \left(1+f'\left(R_0\right)\right) r_{\min }{}^{16 D-2}\epsilon }{4 l^3 \pi  } {dl} \quad \nonumber\\
&+\frac{(8 D-1)\left(4 \left(3-28 D+64 D^2\right) \left(1+f'\left(R_0\right)\right) r_{\min }{}^{16 D-2}-16^d\text{  }\left(-\pi ^{\frac{2}{8-1 D}}\lambda_2\right)^{4 D}\right) \Omega _{D-2} \epsilon }{8\text{  }\pi  r_{\min }{}^3}{dr}_{\min } \quad \nonumber\\
&+O(\epsilon)^2, \label{eq5.20}
\end{align}
where
\begin{align}
\lambda_2=\left(-\frac{(-1)^{-4 D} 2^{1-4 D} \sqrt{1+f'\left(R_0\right)} Q}{ \Omega _{D-2}\text{  }D}\right){}^{\frac{1}{-1+8 D}}. \label{eq5.21}
\end{align}
Using equations (\ref{eq5.20}) and (\ref{eq5.18}),  we have
\begin{align}
&dW_{\min }=\frac{-{  }(16 D-1){  }r_{\min }}{{  }l^2{   }}{dr}_{\min }+\frac{8{  }\sqrt{1+f'\left(R_0\right)}{  }\pi ^{\frac{8D}{8 D-1}}{  }\lambda_2 {dQ}}{(8 D-1) \left(1+f'\left(R_0\right)\right) \Omega _{D-2}r_{\min }{}^{16 D-1} }  \quad \nonumber\\
&~~~~~~~~~- \frac{{  }\left(2 (16 D-3) \left(1+f'\left(R_0\right)\right) r_{\min }{}^{16 D}+16^d r_{\min }{}^2 \left(-\pi ^{\frac{2}{-1+8 D}} \lambda_2\right)^{4 D}\right)}{2{  }\left(1+f'\left(R_0\right)\right){  }{r_{\min }}^{16D+1} }{dr}_{\min }  \quad \nonumber\\
&~~~~~~~~~-\frac{4 \left(2 \sqrt{1+f'\left(R_0\right)} \pi ^{1+\frac{1}{-1+8 D}} Q \mathcal{D}+16^D D \left(-\pi ^{\frac{2}{-1+8 D}} \lambda_2\right)^{4 D} \Omega _{D-2}\right){dQ}}{ (8 D-1) \left(1+f'\left(R_0\right)\right){  }Q \Omega _{D-2}{  }r_{\min }{}^{16D-2} } \quad \nonumber\\
&~~~~~~~~~-\frac{{  }\left(4 \left(3-28 D+64 D^2\right) \left(1+f'\left(R_0\right)\right) {r_{\min }}^{16 D-2}-16^D{  }\left(-\pi ^{\frac{2}{-1+8 D}} \lambda_2\right)^{4 D}\right) \epsilon }{ \left(1+f'\left(R_0\right)\right){  }r_{\min }{}^{16 D}}{dr}_{\min } \quad \nonumber\\
&~~~~~~~~~+\frac{ 2 (16 D-1)r_{\min } \epsilon }{{  }l^3 \Omega _{D-2}}{dl}-\frac{ {l2} \left(1-24 D+128 D^2\right) \epsilon }{{  }l^3{  }}{dr_{\min }}+\mathcal{O}(\epsilon)^2. \label{eq5.22}
\end{align}
In addition, For the extremal black hole, we have $W(r_h) = 0$.  Hence, we can get
\begin{equation}
Q=-\frac{(-1)^{D/4} 2^{\frac{D}{4}-2} D \left(-2^{-\frac{12}{D-2}} \pi ^{-\frac{4}{-2+D}}\lambda _3\right){}^{\frac{1}{4} (D-2)} \Omega _{D-2}}{\sqrt{1+f'\left(R_0\right)}}, \label{eq5.23}
\end{equation}
and
\begin{align}
&{dQ}=-\frac{(-1)^{D/4} 2^{\frac{D}{4}-2} (D-2) \left(-2^{-\frac{12}{-2+D}} \pi ^{-\frac{4}{-2+D}} \lambda _3\right){}^{\frac{1}{4} (-2+D)} }{\sqrt{1+f'\left(R_0\right)} l{ } r_{\min } \left((D-3) l^2+(D-1)r_{\min }{}^2\right)} \quad \nonumber\\
&~~~~~~~~\times \frac{ \left(l \left(\left(6-5 D+D^2\right) l^2+(D-1){dr}_{\min }{}^2\right){dr}_{\min }-2 (D-1)\text{  }r^3{dl}\right)\Omega _{D-2}}{\sqrt{1+f'\left(R_0\right)} l{ } {r}_{\min } \left((D-3) l^2+(D-1)r_{\min }{}^2\right)}, \label{eq5.24}
\end{align}
where
\begin{align}
\lambda _3=\left(-\frac{2^{1-\frac{D}{4}} \left(1+f'\left(R_0\right)\right) r_{\min }{}^{D-2} \left((D-3) l^2+(D-1)r_{\min }{}^2\right)}{l^2}\right){}^{4/D}. \label{eq5.25}
\end{align}
With the help of equations (\ref{eq5.22}), (\ref{eq5.23}) and (\ref{eq5.24}), and the condition $D=4p, p\in \mathbb{N}$. We finally  get
\begin{align}
{dW}_{\min }=\mathcal{O}(\epsilon )^2.   \label{eq5.26}
\end{align}
In the extended phase space, the minimum value of the near-extremal black hole is
\begin{align}
W_{\min }+{dW}_{\min }=\delta _{\epsilon }+\mathcal{O}(\epsilon )^2.  \label{eq5.27}
\end{align}
Obviously, when we considered the condition $\delta_{\epsilon}\rightarrow0$, $\epsilon\rightarrow0$ for equation  (\ref{eq5.27}), we can get the expression $W_{\min}+dW_{\min}=0$, which is reduced to  the extremal case in equation (\ref{eq5.19}).
For  the near-extremal black hole, to determine  the final states precisely, we also  perform higher-order expansion, which is
\begin{align}
&\text{d$\mathcal{W}$}_{\min }=-\frac{\left(2-3 D+D^2\right)\text{  }\epsilon ^2}{l^3}{dl} \quad \nonumber\\
&~~~~~~~~~~~~-\frac{\left( \left(\left(6-5 D+D^2\right) l^2+(D-1) D r^2\right)l {dr}_{\min }-2 (D-1) r_{\min }{}^3\text{dl}\right)\lambda _4\epsilon ^2}{2 \left(1+f'\left(R_0\right)\right) r_{\min }{}^{3+D}l^3\left((D-3) l^2+(D-1) r_{\min }{}^2\right)} \quad \nonumber\\
&~~~~~~~~~~~~+\frac{\text{  }\left(11 D-6 D^2+D^3-6\right)\text{  }\epsilon ^2{dr}_{\min }}{2\text{  }r_{\min }l^2}+\frac{ \left(47 D-12 D^2+D^3-60\right)\text{  }\epsilon ^2{dr}_{\min }}{2\text{  }r_{\min }{}^3} \quad \nonumber\\
&~~~~~~~~~~~~+\frac{3 \lambda _4\epsilon ^2{dr}_{\min }}{2 \left(1+f'\left(R_0\right)\right) r_{\min }{}^{3+D}l^2}+\mathcal{O}(\epsilon )^3, \label{eq5.28}
\end{align}
where
\begin{align}
\lambda _4= l^2r_{\min }{}^2(-1)^{\frac{D}{4}} 2^{\frac{D (D+10)}{4 (D-2)}}\text{  }\pi ^{\frac{D}{D-2}}\text{  }\left(-2^{\frac{12}{2-D}} \pi ^{\frac{4}{2-D}} \lambda _3\right){}^{\frac{D}{4} }. \label{eq5.29}
\end{align}
In this case, we can use equation (\ref{eq5.12}) and equation (\ref{eq5.28}) to define
\begin{align}
\mathcal{W}_E=\frac{\delta _{\epsilon }+\mathcal{O}(\epsilon )^2}{\epsilon ^2}. \label{eq5.30}
\end{align}
Now, in order to visually represent the positive and negative conditions of $\mathcal{W}_E$, we plot Figure 4  for different values of $p$. In these figures, we find that the result is nothing but interesting.
\begin{figure}[htb] \centering\subfigure[{$dl=0.1$ and $p=1$.}] {\includegraphics[scale=0.55,keepaspectratio]{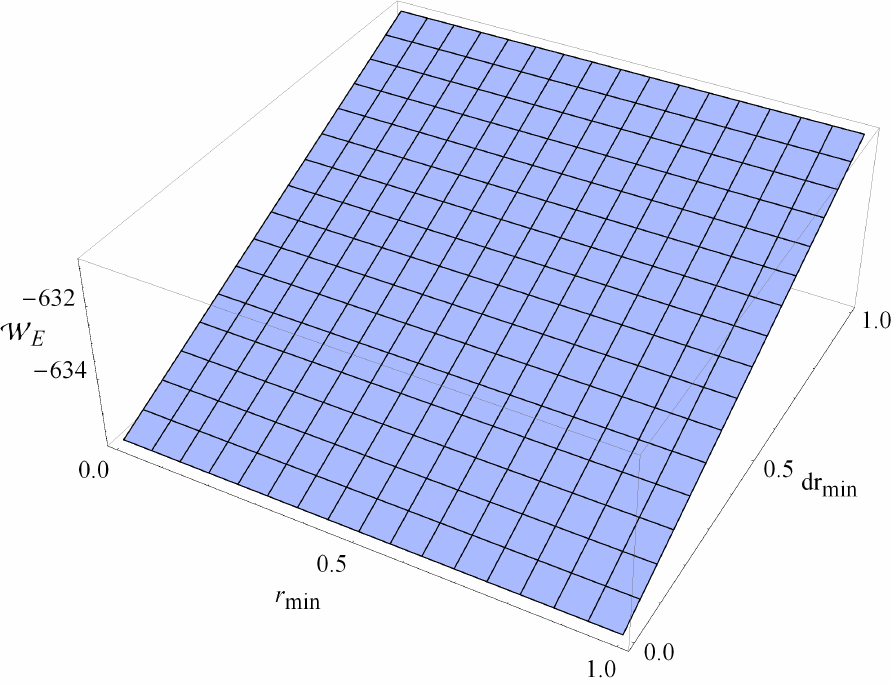}} \quad \centering\subfigure[{$dl=0.1$ and $p=3$.}] {\includegraphics[scale=0.55,keepaspectratio]{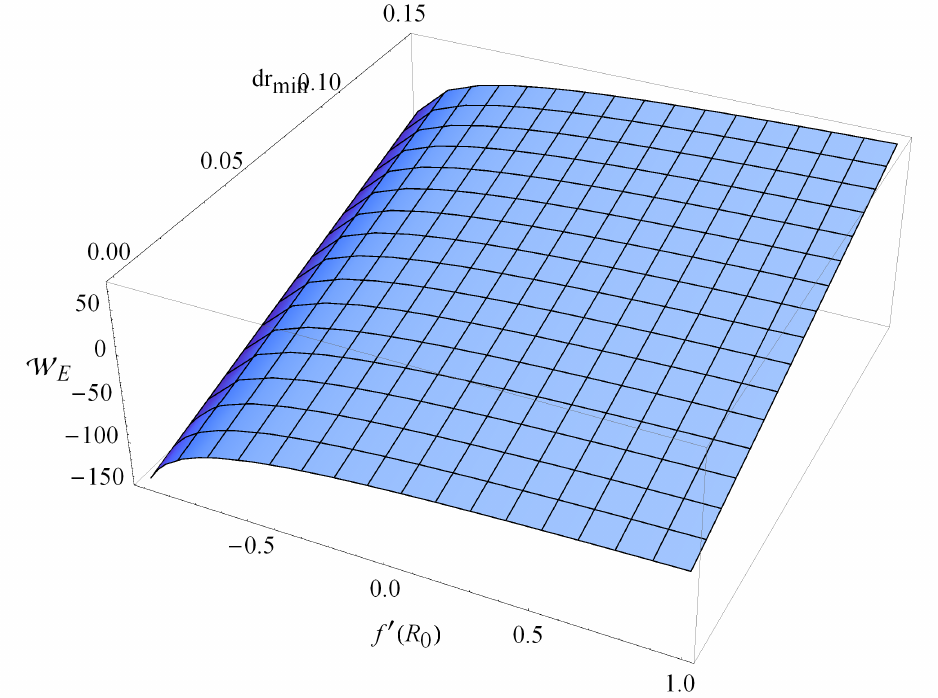}} \quad \centering\subfigure[{$dl=0.1$ and $p=5$.}] {\includegraphics[scale=0.55,keepaspectratio]{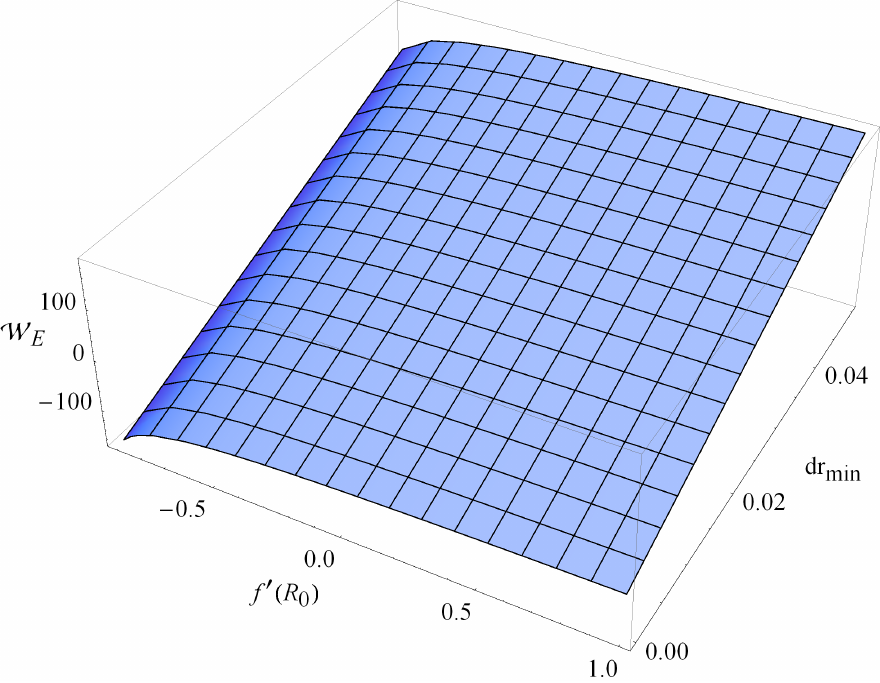}}\\
\caption{The value of $\mathcal{W}_E$ for $Q = 2$, $l=1$, $\Omega _{D-2}= 1$.}
\label{fig:4}
\end{figure}
When $p=1$, which means the four dimensions $f(R)$ black hole, there is no parameter $f'(R_0)$ in the final result of  $\mathcal{W}_E$. It can  be seen clearly that there  is always $\mathcal{W}_E<0$, which is shown in Figure 4 (a). Nevertheless, for the case of $p>1$, that is,  higher dimensional $f(R)$ black holes, the parameter $f'(R_0)$ makes a contribution to the final result $\mathcal{W}_E$  which  leads to $\mathcal{W}_E$ may be positive in the final state, which is shown in Figure 4 (b) and (c).  In this case, there is not a horizon to cover the singularity and the weak cosmic censorship conjecture is violated in the extended phase space. In addition, our result demonstrate that the magnitudes of the violation is different for  the value of parameter $f'(R_0)$, $p$, $dr_{\min}$. In general terms, the configuration of $\mathcal{W}_E$ is different for different values of these parameters, and the magnitudes of the violation is also related to those of the parameters.

\section{Discussion and conclusions}\label{sec:5}
In this paper, we obtained the energy-momentum relation as the charged particle dropped into the higher dimensional charged $f(R)$ black holes by using the Hamilton-Jacobi equation. Based on this relationship, we have verified the thermodynamic laws of black holes under charged particle absorption. In addition, we further examined the validity of the weak cosmic censorship conjecture in the higher-dimensional $f(R)$ AdS black holes.

In the normal phase space,  we found that the first law of thermodynamics was valid  when the charged particle dropped into the higher-dimensional $f(R)$ AdS black holes. Additionally, for the second  law of thermodynamics, the result shows that the variation of the entropy  always increased whether it is extremal or non-extremal black holes, which means the second law of thermodynamics is valid in the normal phase space. According to a more accurate calculate of the shift of the metric function $W(r_{\min})$ under charged particle absorption, the result shows that the final configuration of the black hole does not change, when the extremal $f(R)$  black hole absorbed  the charged particle. In other words, the extremal higher-dimensional $f(R)$ AdS black holes can not be overcharged in the course of the absorption, and the  event horizon of the black hole still holds. In addition, for the case of near-extremal black hole, the minimum value is still negative under charged particle absorption. That is, in both cases, the weak cosmic censorship conjecture are all valid.

In the extended phase space, when the cosmological parameter  is identified as  a variable which is interpreted as a pressure, the results  of thermodynamic laws and weak cosmic censorship conjecture are fairly different from that obtained in the normal phase space. In this case, we find that the first law of thermodynamics is valid under charged particle absorption. However, the results show that the second law of thermodynamics is invalid for extremal and near-extremal black holes. The thermodynamic properties of a black hole, such as the Hawking temperature, Bekenstein-Hawking entropy, and thermodynamic potentials, are all defined on its horizon, especially, the horizon area of the black hole  is proportional to the Bekenstein-Hawking entropy, which means the thermodynamics of a black hole are strongly dependent on the stability of its horizon. Studying  the stability of the horizon is  necessary for the validity of the weak cosmic censorship conjecture, and this conjecture was originally proposed for a stable horizon to prevent the breakdown of the causality at a naked singularity. Thus, we need  to further prove the validity of the conjecture when  the second law appears to be violated with the pressure term.  Therefore, we judged the existences of the event horizon by evaluating the minimum value of the function $W(r)$. In this paper, our results show that the function $W(r)$ does not also change for the extremal black hole. That is,  extremal higher-dimensional $f(R)$ AdS black holes can not be destroyed  in the course of the absorption process, and the weak cosmic censorship conjecture is still valid  in the extended phase space. Interestingly, for the near-extremal higher-dimensional $f(R)$ black holes,  the shift of the minimum value is quite different from that in the case without the pressure term. Different with Ref. \cite{ref41}, the effect of the second-order small $\mathcal O(\epsilon^2 )$ to the final result is presented in our calculation, where the figures of the relation between $\delta _{\epsilon }$ and $\mathcal O(\epsilon^2 )$ is plotted. In this case,  the result shows that there is still $W(r_{\min}+dr_{\min})<0$ in the case of $p=1$,  where the parameter $f'(R)$ does not makes a contribution to $W(r_{\min}+dr_{\min})$.  That is, the weak cosmic censorship conjecture is valid in the  4-dimensional near-extremal $f(R)$ black holes. However, for the  higher-dimensional $f(R)$ AdS black hole where $p>1$, our results show that $W(r_{\min}+dr_{\min})>0$, and we find that parameter $f'(R)$  makes a contribution to $W(r_{\min}+dr_{\min})$. It is worth noting that this result is quite different from that without parameters $f'(R)$ in the final state. In other words, the weak cosmic censorship conjecture may be invalid  when the charged particle dropped into the higher-dimensional near-extremal $f(R)$ black holes. In a conclusion, it implies that the violations of the cosmic censorship conjecture depending on the parameter $f'(R_0)$, and the magnitudes of those violations are relevant to those of the parameters. Therefore, the parameter $f'(R_0)$ plays a very important role, and its effect to the weak cosmic censorship conjecture cannot be neglected.

\vspace{10pt}

\noindent {\bf Acknowledgments}

\noindent
This work is supported  by the National
Natural Science Foundation of China (Grant Nos. 11875095, 11903025), and Basic Research Project of Science and Technology Committee of Chongqing (Grant No. cstc2018jcyjA2480).

\end{document}